\documentclass[11pt,twoside]{article} 
\usepackage{asp2004}
\usepackage{epsf}
\usepackage{psfig}
\usepackage{lscape}

\begin{document} 
\title{Comparison of White Dwarf Mass Determinations}
\author{S. Boudreault and P. Bergeron} 
\affil{D\'epartement de Physique, Universit\'e de Montr\'eal, C.P. 6128, Succursale Centre-Ville, Montr\'eal, Qu\'ebec, Canada, H3C 3J7}
\begin{abstract}
White dwarf masses can be determined in several ways. Here we compare masses obtained from Balmer line spectroscopy with those derived from trigonometric parallax measurements for an ensemble of cool hydrogen-atmosphere white dwarfs.
\end{abstract}
\section{Fitting Technique}
Our photometric and spectroscopic sample is composed of all DA stars in the trigonometric parallax sample of Bergeron, Leggett, \& Ruiz (2001; BLR hereafter). We restrict, however, our analysis to objects above 6500 K (52 stars), for which the spectroscopic method of determining atmospheric parameters remains reliable. The sample is first analyzed by fitting the optical $BVRI$ and infrared $JHK$ photometry using the method described in BLR and references therein. Only $T_{\rm eff}$ and the solid angle $\pi(R/D)^{2}$ are considered free parameters, where $(R/D)$ is the ratio of the radius of the star to its distance from Earth. The distances are obtained from the trigonometric parallax measurements, and the resulting radii are converted into masses using the evolutionary models of Fontaine, et al. (2001) with thick hydrogen layers. Spectroscopic observations have been secured for the DA stars in our sample using the Steward 2.3-m telescope equipped with the Boller \& Chivens spectrograph. The high Balmer line spectra are fitted with a grid of model atmospheres following the procedure described in Bergeron, et al. (1992). The resulting $T_{\rm eff}$ and log \textit{g} values are then converted into masses using the same evolutionary models as above. 
\begin{figure}[!ht]
\plotfiddle{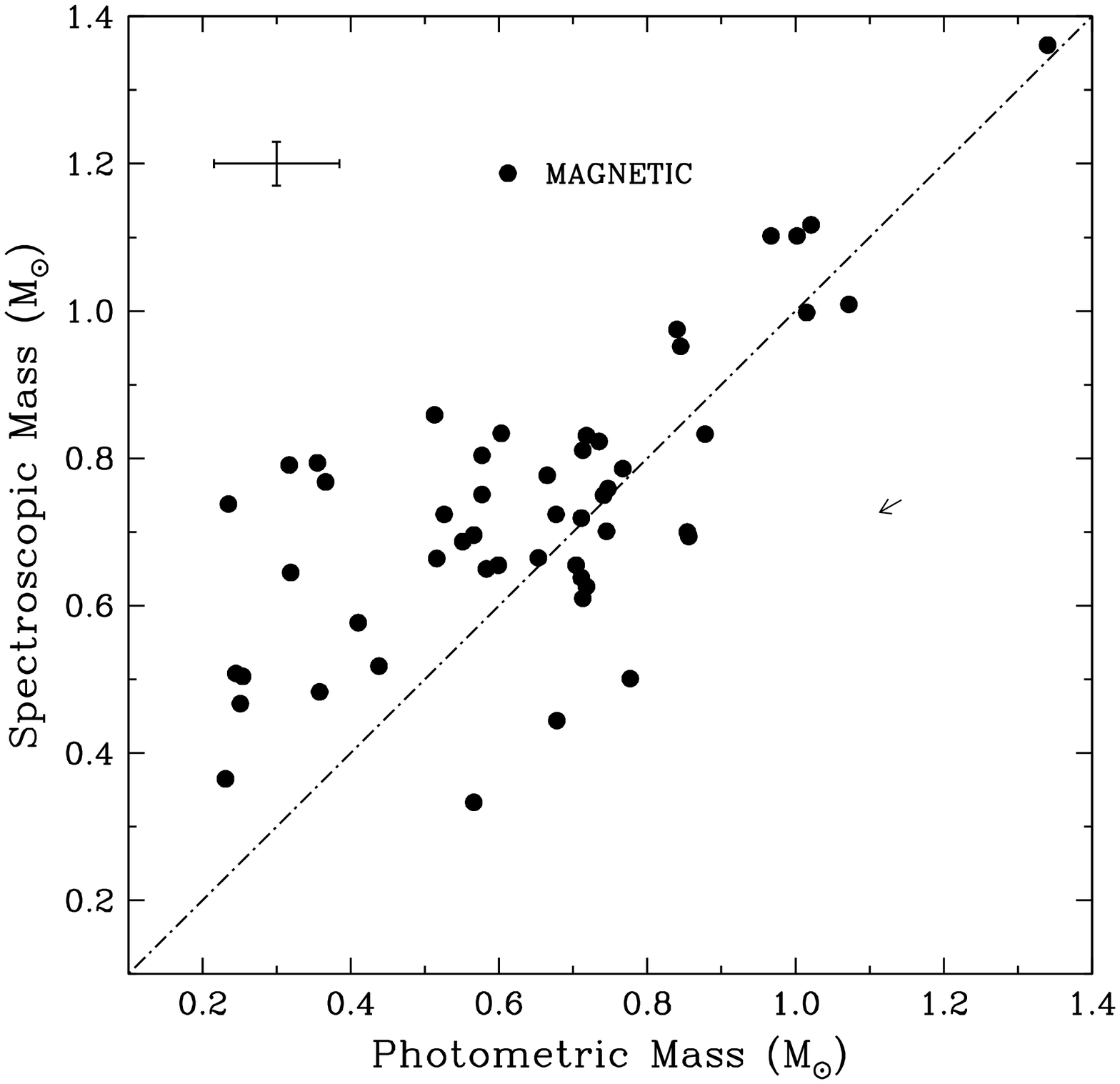}{4.2 cm}{0}{30}{30}{-5 pt}{-45 pt}
\plotfiddle{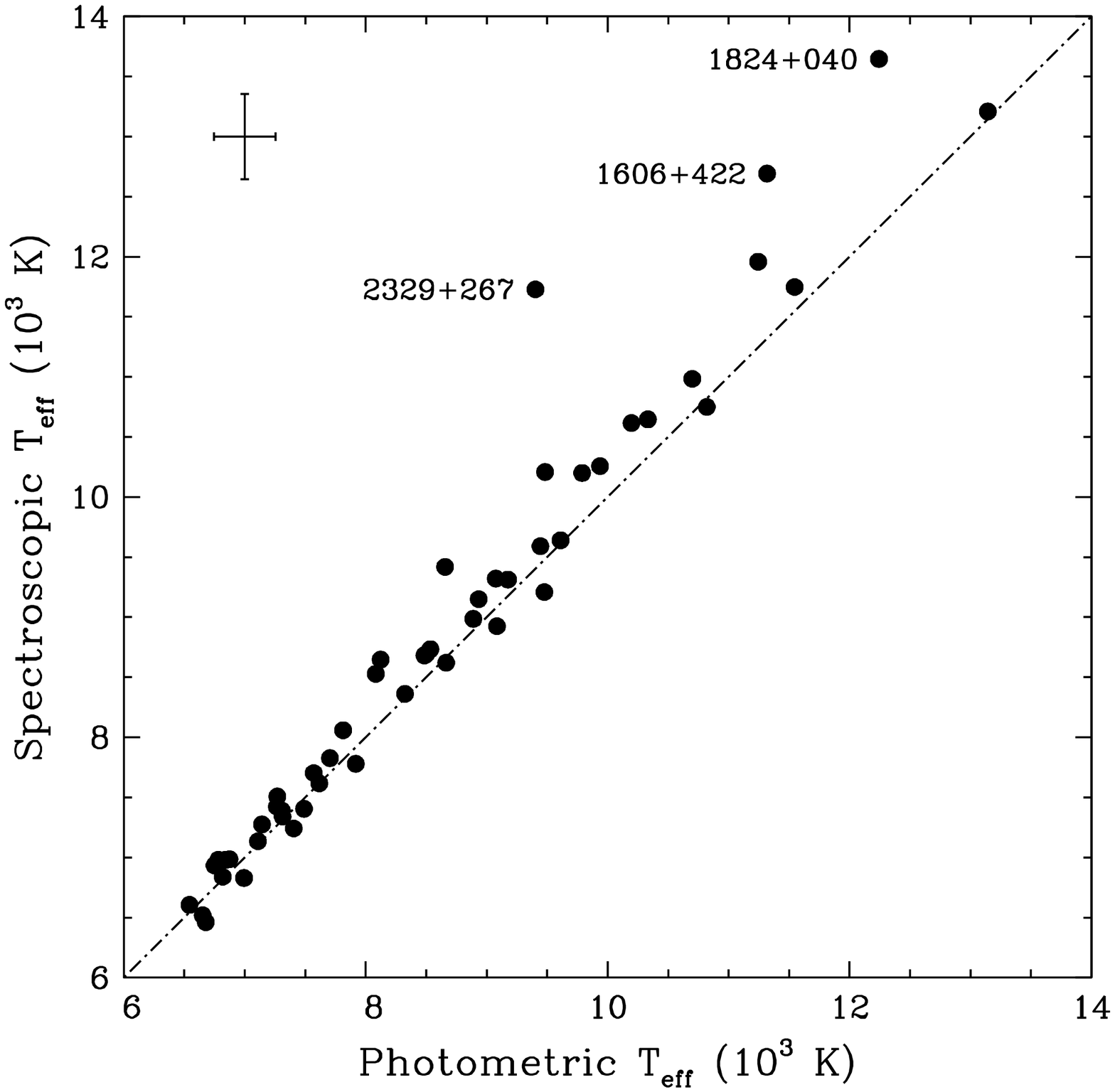}{0.0 cm}{0}{30}{30}{-195 pt}{-20 pt}
\caption{Comparison of the spectroscopic and photometric temperature (\textit{left panel}) and masses (\textit{right panels}) for our 52 white dwarfs.}
\end{figure}
\section{Comparison of Spectroscopic and Photometric Stellar Parameters}
We compare in Figure 1 (left panel) the spectroscopic and photometric temperatures for all 52 stars in our sample. The agreement between the spectroscopic and photometric temperatures for our sample is excellent, with the exception of WD 2329+267, a magnetic white dwarf whose line profiles exhibit strong Zeeman splitting, WD 1606+422, a suspected double degenerate (BLR), and WD 1824+040, a confirmed double degenerate (Maxted \& Marsh; 1999). The comparison of spectroscopic and photometric masses for our 52 objects is displayed in Figure 1 (right panel). The vector shown in the figure indicates how the masses would be affected if thin hydrogen layer models were used instead of thick hydrogen models to derive masses. For most objects the mass determinations yield significantly different results, as can be seen by the size of the uncertainties (cross in upper left corner). In general, the spectroscopic masses are larger than those inferred from photometry and trigonometric parallaxes. One obvious exception is LHS 4033 near 1.35 $M_\odot$, for which both estimates agree perfectly (see also Dahn \textit{et al.}; 2004). Interestingly enough, there appears to be a large number of low mass ($M$ $<$ 0.5 $M_\odot$) stars in the photometric mass distribution that is not observed in the spectroscopic distribution. This suggests that the low mass component may be contaminated by unresolved double degenerates, which would appear as overluminous, and thus interpreted as low mass white dwarfs when analyzed with the photometric technique.
\begin{figure}[!ht]
\plotfiddle{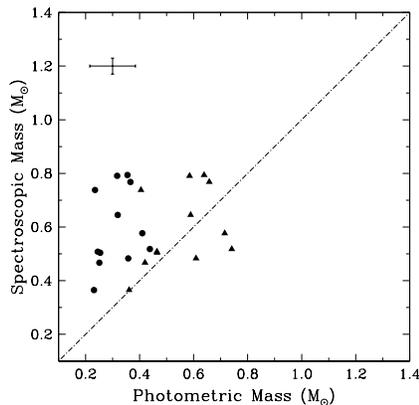}{4.4 cm}{0}{27}{27}{-85 pt}{-25 pt}
\caption{Comparison of the spectroscopic and photometric masses for the 13 objects below $M_{\rm phot}$ = 0.48 $M_\odot$ (\textit{circles}) by assuming that two white dwarfs of equal temperatures and radii contribute to the total luminosity (\textit{triangles}).}
\end{figure}
\section{Unresolved Double Degenerates}
If we redo our analysis for the 13 objects below $M_{\rm phot}$ $=$ 0.48 $M_\odot$ in Figure 1 by assuming that two white dwarfs of equal temperatures and radii contribute to the total luminosity, we obtain the results shown in Figure 2. Of course, the spectroscopic mass determinations are not affected since both stars are assumed to be identical. We see that in most cases, but not all, it is possible to reconcile the spectroscopic and photometric masses. One of the objects in Figure 2 is L870-2, a well-known double degenerate with almost identical atmospheric parameters.
\section{The Presence of Helium in the Atmospheres of Cool DA Stars} 
The mass discrepancy for the DA stars in our sample can also be explained in terms of the presence of helium in the atmospheres of cool DA stars, brought to the surface by the hydrogen convection zone, and spectroscopically invisible at these low temperatures. Indeed, as demonstrated by Bergeron et al. (1990), the presence of helium tends to increase the atmospheric pressure, which would be interpreted as resulting from a high surface gravity (or high mass) when analyzed with pure hydrogen models. 
\begin{figure}[!ht]
\plotfiddle{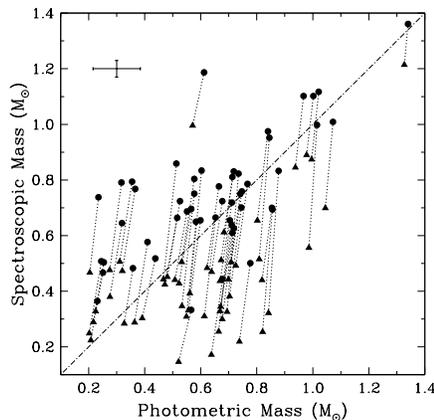}{4.7 cm}{0}{28}{28}{-90 pt}{-25 pt}
\caption{Comparison of the spectroscopic and photometric masses assuming a pure hydrogen atmosphere (\textit{circles}) and a mixed composition of  $N$(He)$/$$N$(H) $=$ 1 (\textit{triangles}). }
\end{figure}
To illustrate more quantitatively the effect, we compare in Figure 3 the spectroscopic and photometric masses obtained under the assumption of pure hydrogen atmospheres, as well as with a mixed composition of $N$(He)$/$$N$(H) $=$ 1. While the photometric masses remain relatively unchanged, the spectroscopic masses are significantly reduced. Clearly, for all objects, with the exception of some of the low mass stars, it is possible to reconcile the spectroscopic and photometric masses by adjusting the individual helium abundance of each star. 
\begin{figure}[!ht]
\plotfiddle{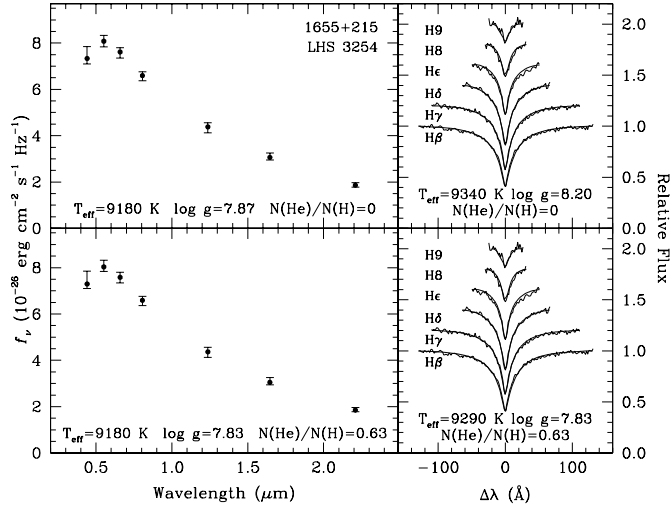}{6.0 cm}{0}{44}{44}{-130 pt}{-100 pt}
\caption{Comparison of the spectroscopic and photometric masses assuming a pure hydrogen layer (\textit{top panels}) and a mixed composition of $N$(He)$/$$N$(H) $=$ 0.63 (\textit{bottom panels}). }
\end{figure}
In the top panels of Figure 4, we compare the photometric and spectroscopic solutions for LHS 3254 (WD 1655+215) under the assumption of a pure hydrogen atmospheric composition. In the bottom panels, the helium abundance has been adjusted to ensure that the photometric and spectroscopic log \textit{g} values (or masses) agree. We see that in this star, a helium abundance of $N$(He)$/$$N$(H) $=$ 0.63 is required to match the log \textit{g} (or mass) determinations. Also, the photometric and spectroscopic temperatures agree even better when helium is included. Finally, the spectroscopic mass of 0.50 $M_\odot$ inferred from the mixed He/H composition is in much better agreement with those of hotter sibblings, in contrast with the pure hydrogen spectroscopic solution of 0.72 $M_\odot$. Helium abundances for individual objects will be presented elsewhere.
\section{Conclusion}
We thus conclude that the existence of double degenerates, or the presence of helium in the atmosphere of cool DA stars, or both, are responsible for the spectroscopic and photometric mass discrepancies observed in Figure 1. The presence of helium in cool DA stars will reduce the inferred spectroscopic masses and may help resolve part of the problem with larger-than-average masses observed in spectroscopic analyses of DA stars at low temperatures. 
\acknowledgements{This work was supported in part by the NSERC Canada and by the Fund FQRNT (Qu\'ebec).}

\end{document}